\documentstyle[12pt,psfig]{article}

\catcode`\@=11
\long\def\@makefntext#1{
\protect\noindent \hbox to 3.2pt {\hskip-.9pt  
$^{{\ninerm\@thefnmark}}$\hfil}#1\hfill}                

\def\@makefnmark{\hbox to 0pt{$^{\@thefnmark}$\hss}}  
        
\def\ps@myheadings{\let\@mkboth\@gobbletwo
\def\@oddhead{\hbox{}
\rightmark\hfil\ninerm\thepage}   
\def\@oddfoot{}\def\@evenhead{\ninerm\thepage\hfil
\leftmark\hbox{}}\def\@evenfoot{}
\def\sectionmark##1{}\def\subsectionmark##1{}}

\renewcommand{\thefootnote}{\fnsymbol{footnote}}

\newcounter{sectionc}\newcounter{subsectionc}\newcounter{subsubsectionc}
\renewcommand{\section}[1] {\vspace*{0.6cm}\addtocounter{sectionc}{1} 
\setcounter{subsectionc}{0}\setcounter{subsubsectionc}{0}\noindent 
        {\normalsize\bf\thesectionc. #1}\par\vspace*{0.4cm}}
\renewcommand{\subsection}[1] {\vspace*{0.6cm}\addtocounter{subsectionc}{1} 
        \setcounter{subsubsectionc}{0}\noindent 
        {\normalsize\it\thesectionc.\thesubsectionc. #1}\par\vspace*{0.4cm}}
\renewcommand{\subsubsection}[1]
{\vspace*{0.6cm}\addtocounter{subsubsectionc}{1}
        \noindent {\normalsize\rm\thesectionc.\thesubsectionc.\thesubsubsectionc. 
        #1}\par\vspace*{0.4cm}}

\newcounter{appendixc}
\newcounter{subappendixc}[appendixc]
\newcounter{subsubappendixc}[subappendixc]

\renewcommand{\appendix}[1] {\vspace*{0.6cm}
        \refstepcounter{appendixc}
        \setcounter{figure}{0}
        \setcounter{table}{0}
        \setcounter{equation}{0}
        \renewcommand{\thefigure}{\Alph{appendixc}.\arabic{figure}}
        \renewcommand{\thetable}{\Alph{appendixc}.\arabic{table}}
        \renewcommand{\theappendixc}{\Alph{appendixc}}
        \renewcommand{\theequation}{\Alph{appendixc}.\arabic{equation}}
        \noindent{\bf Appendix \theappendixc #1}\par\vspace*{0.4cm}}

\def\abstracts#1{{
        \centering{\begin{minipage}{12.2truecm}\footnotesize\baselineskip=12pt\noindent
        \centerline{\footnotesize ABSTRACT}\vspace*{0.3cm}
        \parindent=0pt #1
        \end{minipage}}\par}} 

\renewenvironment{thebibliography}[1]
        {\begin{list}{\arabic{enumi}.}
        {\usecounter{enumi}\setlength{\parsep}{0pt}
\setlength{\leftmargin 1.25cm}{\rightmargin 0pt}
         \setlength{\itemsep}{0pt} \settowidth
        {\labelwidth}{#1.}\sloppy}}{\end{list}}

\newcounter{itemlistc}
\newcounter{romanlistc}
\newcounter{alphlistc}
\newcounter{arabiclistc}

\newcommand{\fcaption}[1]{
        \refstepcounter{figure}
        \setbox\@tempboxa = \hbox{\footnotesize Fig.~\thefigure. #1}
        \ifdim \wd\@tempboxa > 6in
           {\begin{center}
        \parbox{6in}{\footnotesize\baselineskip=12pt Fig.~\thefigure. #1}
            \end{center}}
        \else
             {\begin{center}
             {\footnotesize Fig.~\thefigure. #1}
              \end{center}}
        \fi}

\newcommand{\tcaption}[1]{
        \refstepcounter{table}
        \setbox\@tempboxa = \hbox{\footnotesize Table~\thetable. #1}
        \ifdim \wd\@tempboxa > 6in
           {\begin{center}
        \parbox{6in}{\footnotesize\baselineskip=12pt Table~\thetable. #1}
            \end{center}}
        \else
             {\begin{center}
             {\footnotesize Table~\thetable. #1}
              \end{center}}
        \fi}

\def\@citex[#1]#2{\if@filesw\immediate\write\@auxout
        {\string\citation{#2}}\fi
\def\@citea{}\@cite{\@for\@citeb:=#2\do
        {\@citea\def\@citea{,}\@ifundefined
        {b@\@citeb}{{\bf ?}\@warning
        {Citation `\@citeb' on page \thepage \space undefined}}
        {\csname b@\@citeb\endcsname}}}{#1}}

\newif\if@cghi
\def\cite{\@cghitrue\@ifnextchar [{\@tempswatrue
        \@citex}{\@tempswafalse\@citex[]}}
\def\citelow{\@cghifalse\@ifnextchar [{\@tempswatrue
        \@citex}{\@tempswafalse\@citex[]}}
\def\@cite#1#2{{$\null^{#1}$\if@tempswa\typeout
        {IJCGA warning: optional citation argument 
        ignored: `#2'} \fi}}

 1
 1
 1

\font\ninerm=cmr9

\def\app#1#2#3{          {\it Astroparticle Phys. }{\bf #1} (19#2) #3}

\def\nps#1#2#3{          {\it Nucl.~Phys.~B (Proc.Suppl.) }
                         {\bf #1} (19#2) #3} 
\def\np#1#2#3{           {\it Nucl.~Phys. }{\bf #1} (19#2) #3}
\def\pl#1#2#3{           {\it Phys.~Lett. }{\bf #1} (19#2) #3}
\def\pr#1#2#3{        {\it Phys.~Rev. }{\bf #1} (19#2) #3}

\def\n.c.#1#2#3{         {\it Nuovo Cim. }{\bf #1} (19#2) #3}
\def\r.n.c.#1#2#3{       {\it Riv.~del Nuovo Cim. }{\bf #1} (19#2) #3}

\def\yf#1#2#3{           {\it Yad.~Fiz. }{\bf #1} (19#2) #3}

\def\jetp#1#2#3{         {\it Sov.~Phys.~JETP }{\bf #1} (19#2) #3}
\def\jetpl#1#2#3{         {\it JETP Lett. }{\bf #1} (19#2) #3}

\def\ppnp#1#2#3{           {\it Prog.~Part.~Nucl.~Phys. }{\bf #1} (19#2) #3}

\def\eq#1{{eq. (\ref{#1})}}

\def\gsim{\;\raise0.3ex\hbox{$>$\kern-0.75em\raise-1.1ex\hbox{$\sim$}}\;}
\def\lsim{\;\raise0.3ex\hbox{$<$\kern-0.75em\raise-1.1ex\hbox{$\sim$}}\;}
\newcommand{\Frac}[2]{\frac{\displaystyle #1}{\displaystyle #2}}
\begin{document}

\centerline{\normalsize\bf NEW TESTS FOR NEUTRINOS IN}
\baselineskip=22pt
\centerline{\normalsize\bf  LOW-ENERGY SOLAR EXPERIMENTS}

\centerline{\footnotesize SERGIO PASTOR\footnote{Work in collaboration
with J.~Segura (U.~Elche), V.B.~Semikoz (IZMIRAN) and J.W.F.~Valle
(IFIC)}}
\baselineskip=13pt
\centerline{\footnotesize\it Departament de F\'{\i}sica Te\`orica, 
IFIC-Universitat de Val\`encia, 46100 Burjassot, Val\`encia, SPAIN}
\baselineskip=12pt
\centerline{\footnotesize E-mail: sergio@flamenco.ific.uv.es}

\vspace*{0.9cm} \abstracts{We show how future solar neutrino
  experiments in the low energy region can be used to test novel
  neutrino properties. Information on the Majorana nature or neutrino
  magnetic moments can be extracted from the observation of electron
  anti-neutrinos from the Sun and the measurement of an azimuthal
  asymmetry in the total number of events, respectively.}

\normalsize\baselineskip=15pt
\setcounter{footnote}{0}
\renewcommand{\thefootnote}{\alph{footnote}}
\section{Low-energy electron Anti-neutrinos from the Sun}

Finding a signature for the Majorana nature of neutrinos or,
equivalently, for the violation of lepton number in Nature is a
fundamental challenge in particle physics \cite{fae}. All attempts for
distinguishing Dirac from Majorana neutrinos {\em directly} in
laboratory experiments have proven to be a hopeless task, due to the
V-A character of the weak interaction, which implies that all such
effects vanish as the neutrino mass goes to zero. We suggest an
alternative way in which one might probe for the possibility of
$L$-violation which is not {\em directly} induced by the presence of a
Majorana mass. Although, Majorana masses will be required at some
level, but the quantity which is directly involved is the transition
amplitude for $\nu_e \rightarrow \bar{\nu}_e$ conversions
in the Sun.

Here we propose to probe for the possible existence of
$L$-violating processes in the solar interior that can produce an
$\bar{\nu}_e$ component in the neutrino flux.  The idea is that, even
though the nuclear reactions that occur in a normal star like our Sun
do not produce directly right-handed active neutrinos ($\bar{\nu}_a$)
these may be produced by combining the chirality-flipping transition
$\nu_{eL} \rightarrow \bar{\nu}_{a R}$ with the standard
chirality-preserving MSW conversions $\nu_{eL} \rightarrow \nu_{\mu
L}$ through cascade conversions like $\nu_{eL} \rightarrow
\bar{\nu}_{\mu R} \rightarrow \bar{\nu}_{eR}$ or $\nu_{eL} \rightarrow
\nu_{\mu L} \rightarrow \bar{\nu}_{eR}$.  These conversions arise as a
result of the interplay of two types of mixing \cite{akhpetsmi}: one
of them, matter-induced flavour mixing, leads to MSW resonant
conversions which preserve the lepton number $L$, whereas the other is
generated by the resonant interaction of a Majorana neutrino
transition magnetic moment with the solar magnetic field
\cite{LAM}. This violates the $L$ symmetry by two units ($\Delta L =
\pm 2$) and is an explicit signature of the Majorana nature of the
neutrino \cite{BFD}.

We consider neutrino-electron scattering in future underground solar
neutrino experiments in the low-energy region, below the threshold for
$\bar{\nu}_e + p \rightarrow n +e^+$, such as HELLAZ \cite{hellaz} or BOREXINO
\cite{borexino}. They will have low energy thresholds ($100$ keV and
$250$ keV, respectively). BOREXINO is designed to take
advantage of the characteristic shape of the electron recoil energy
spectrum from the $^7$Be neutrino line, while 
the HELLAZ experiment is intended to measure the
fundamental $pp$ neutrinos.

The complete expression for the differential cross section of the weak
process $\nu e \rightarrow \nu e$, as a function of the electron
recoil energy $T$, in the massless neutrino limit, can be written as
(see for instance ref.~\cite{Sem97})
\begin{equation}
\label{totdsigma}
\frac{d \sigma}{dT} (\omega,T)= \frac{2 G_F^2 m_e}{\pi} \Bigl [
P_e h(g_{eL},g_R) 
+ P_{\bar{e}} h(g_R,g_{eL})
+ P_a h(g_{aL},g_R)
+ P_{\bar{a}} h(g_R,g_{aL})\Bigr ]
\end{equation}
where
$h(a,b) \equiv  a^2 + b^2 (1- T/\omega)^2 -
ab m_e T/\omega^2$
and $g_{eL} = \sin^2 \theta_W+0.5$ , $g_{a L} = \sin^2 \theta_W-0.5$
($a=\mu,\tau$) and $g_R = \sin^2 \theta_W$ are the weak couplings of
the Standard Model, and $\omega$ is the energy of the incoming
neutrino. The parameter $P_e$ in the equation above is the survival
probability of the initial left-handed electron neutrinos, while
$P_{\bar{e}}$, $P_a$ and $P_{\bar{a}}$ are the appearance
probabilities of the other species, that may arise in the Sun as a
result of the processes $\nu_{eL} \to \bar{\nu}_{eR}$, $\nu_{eL}\to
\nu_{aL}$ or $\nu_{eL} \to \bar{\nu}_{aR}$, respectively. These
parameters obey the unitarity condition $P_e(\omega) +
P_{\bar{e}}(\omega) + P_a(\omega) + P_{\bar{a}}(\omega) = 1$.  In
general they are obtained from the complete $4\times 4$ 
Hamiltonian describing the evolution of the neutrino system
\cite{BFD}. They depend on the neutrino energy $\omega$,
on the solar magnetic field through $\mu_\nu B_\perp$
and on the neutrino mixing parameters $\Delta m^2$, $\sin ^22\theta$.

In the $L$-violating processes one has in general all four
contributions shown in \eq{totdsigma}. In contrast, in the case where
lepton number is conserved (like in MSW conversions), the solar
neutrino flux will consist of {\em neutrinos}, so only the first and
third terms in \eq{totdsigma} contribute.  It follows that the
differential cross section will be {\em different} in the case where
$\nu_e$'s from the Sun get converted to electron $\bar{\nu}_e$'s. Is
it possible to measure this difference in future neutrino experiments?

The relevant quantity to be measured in neutrino scattering
experiments is the energy spectrum of events, namely
\begin{equation}
\label{spectrum}
\frac{dN_\nu}{dT} = N_e \sum_{i} \phi_{0i} 
\int^{\omega_{max}}_{\omega_{min}(T)}
d \omega \lambda_i(\omega) 
\langle \frac{d \sigma}{dT} (\omega,T) \rangle
\end{equation}
where $d \sigma/dT$ is given in \eq{totdsigma} and $N_e$ is the number
of electrons in the fiducial volume of the detector
\footnote{For simplicity we take
the detector efficiency as unity for energies above the
threshold.}.  The sum in the above equation is done over the solar
neutrino spectrum, where $i$ corresponds to the different reactions
$i= pp$, $^7$Be, $pep$, $^8$B $\ldots$, characterized by an integral
flux $\phi_{0i}$ and a differential spectrum $\lambda_i(\omega)$ (for
neutrinos coming from two-body reactions, one has $\lambda_i(\omega) =
\delta(\omega - \omega_i)$). The lower limit for the neutrino energy
is $\omega_{min} (T) = (T + \sqrt{T^2 +2m_eT})/2$, while the upper
limit 
\begin{figure}
\centerline{\protect\hbox{\psfig{file=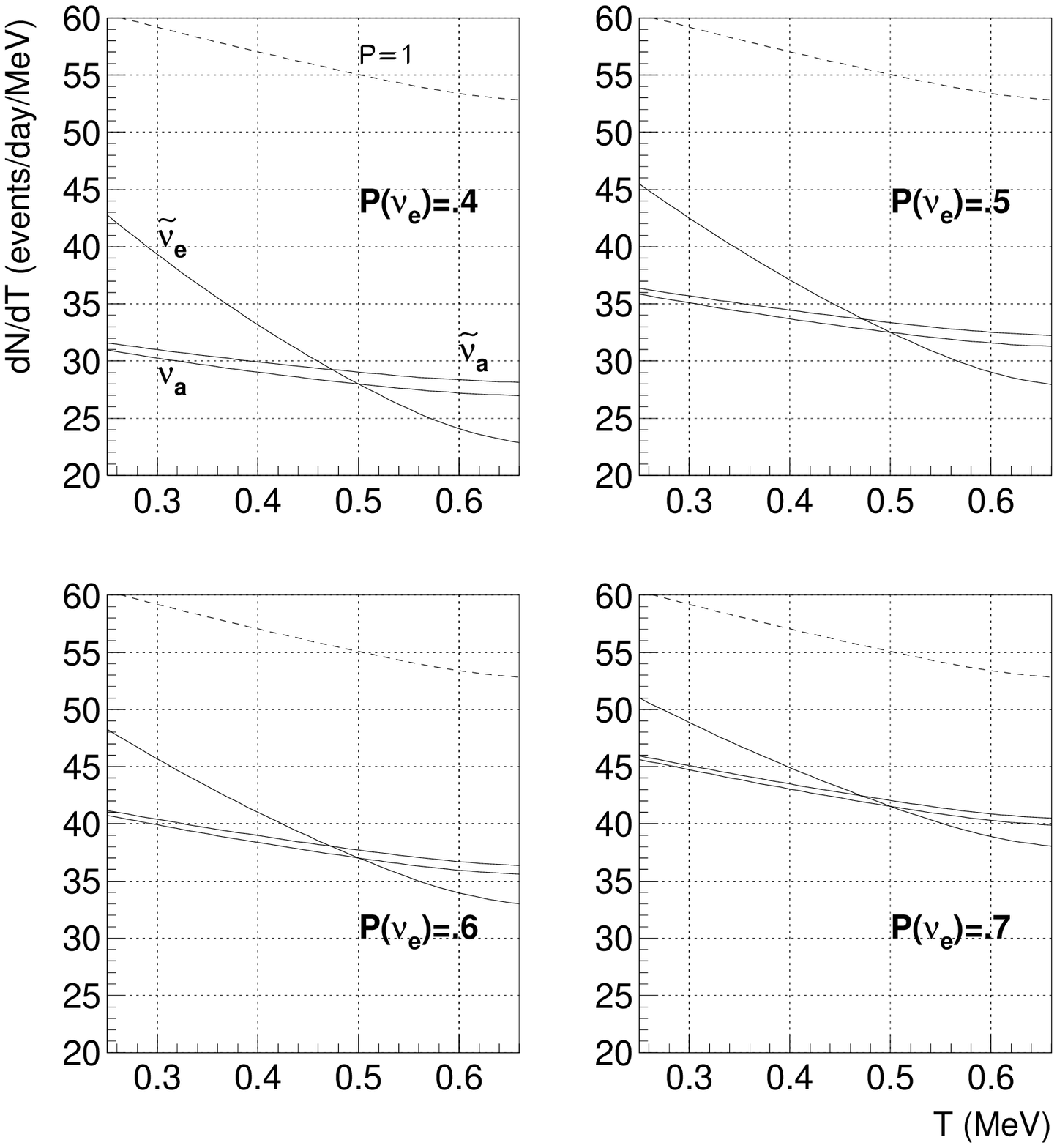,width=0.99\textwidth}}}
\fcaption{Energy spectrum of events corresponding to $^7$Be solar
neutrinos for the BOREXINO experiment.  The upper line corresponds
to the case where one has no neutrino conversions ($P_e=1$). When
electron anti-neutrinos are present in the solar flux the results
are the lines labeled with $\bar{\nu}_e$, calculated for the
indicated value of $P_e$ and the corresponding amount of
$\bar{\nu}_a$. The cases of $\nu_e \rightarrow \nu_{\mu,\tau}$ and
$\nu_e \rightarrow \bar{\nu}_{\mu,\tau}$ are the lower lines with
labels $\nu_a$ and $\bar{\nu}_a$, respectively.}
\label{fig1}
\end{figure}
$\omega_{max}$ corresponds to the maximum neutrino energy. In
order to take into account the finite resolution in the measured
electron recoil energy, we perform a Gaussian average of the cross
section, indicated by $\langle \ldots \rangle$ in \eq{spectrum}.  For
further details, see ref. \cite{paper}.

We have calculated the averaged energy spectrum of events
for the two experiments in the simple case where the
parameters $P_i$ do not depend on the neutrino energy. Here we
present our results for BOREXINO in figure \ref{fig1}, 
where one can see that it is possible to distinguish
the case with $\bar{\nu}_e$ considering the behaviour of the cross
section for low energies. It is the {\em slope} of the measured
spectrum the key for recognizing the presence of $\bar{\nu}_e$'s
in the solar neutrino flux, and correspondingly the
presence of $L$-violating processes which can only exist if neutrinos
are Majorana particles.

The shortcoming of the above discussion is that 
we have neglected the energy dependence of the
physical parameters $P_i$. One must calculate
the averaged $\nu-e$ cross section using analytical expressions for
$P_i=P_i(\omega)$.  However, since the $^7$Be
neutrinos are mono-energetic, whatever the mechanism that produces the
deficit is, their survival probability will take on a constant value
$P_e(\omega_{Be})$. Therefore one can apply directly the results we
have obtained for constant $P_i$ for the range of electron recoil
energy where the contribution of $^7$Be neutrinos dominates. In
ref. \cite{paper} we give a discussion of the experimental uncertainties.

There are however stringent bounds on the presence of solar
$\bar{\nu}_e$'s in the high energy region ($^8$B). These would
interact within the detector through the process $\bar{\nu}_e + p
\rightarrow n +e^+$.  This process, which has an energy threshold of
$E_\nu = m_n -m_p +m_e \simeq 1.8$ MeV, has not been found to occur in
the Kamiokande experiment nor in the very recent data from
Super-Kamiokande \cite{fiorentini97}.  Also the results from LSD are
negative \cite{aglietta96}.

As we show in ref. \cite{paper} for the specific scenario presented in
ref. \cite{akhpetsmi}, the co-existence of a suppressed production of
high-energy $\bar{\nu}_e$'s and a sizeable flux of anti-neutrinos at
energies below $1.8$ MeV can be easily understood theoretically. The
resonance in the $\nu_{eL} \rightarrow \bar{\nu}_{eR}$ conversions
can lie in the energy region below $1$ MeV (relevant for HELLAZ or
BOREXINO) provided that the neutrino parameters have reasonable
values, so that the conversion probability is small for energies
$\omega \gg 1$ MeV. Therefore the anti-neutrino flux would be {\em
hidden} in the background and therefore unobservable in
Super-Kamiokande.
Our conclusion is that neutrino conversions within the Sun can result
in partial polarization of the initial fluxes, in such a way as to
produce a sizeable $\bar{\nu}_e$ component without conflicting present
Super-Kamiokande data. The observation of $\bar{\nu}_e$'s from the Sun
in future neutrino experiments in the low energy region could lead to
the conclusion that the neutrinos are Majorana particles.

\section{Neutrino magnetic moments and low-energy
solar neutrino-electron scattering experiments}

In this section we are concerned with a particular effect in
neutrino-electron scattering for solar neutrinos which possess
a Dirac magnetic moment or Majorana transition magnetic moments,
that we will note generically as $\mu_\nu$. The
latter is especially interesting first of all because Majorana
neutrinos are more fundamental and they arise in most models
of particle physics beyond the standard model \cite{fae}, and because
its effects can be resonantly enhanced in matter \cite{LAM}, providing
one of the most attractive solutions to the solar neutrino
problem \cite{revAkhmedov}. Another practical advantage in favour of
Majorana transition moments is that, in contrast to Dirac-type
$\mu_\nu$'s, these are substantially less stringently constrained
by astrophysics \cite{RaffeltA}.

For {\em longitudinally polarized neutrinos} the weak interaction and
the electromagnetic interaction amplitudes on electrons do not
interfere, since the weak interaction preserves neutrino helicity
while the electromagnetic does not. As a result the cross section
depends quadratically on $\mu_\nu$.  However if there exists a process
capable of converting part of the initially fully polarized $\nu_e$'s,
then an {\em interference term} arises proportional to $\mu_\nu$, as
pointed out e.g. in ref. \cite{Barbieri}.  This term depends on the
angle between the component of the neutrino spin transverse to its
momentum and the momentum of the outgoing recoil electron.  Therefore
the number of events measured in an experiment exhibits an {\em
  asymmetry} with respect to the above defined angle.  The asymmetry
will not show up in terrestrial experiments even with stronger
magnetic fields, since only in the Sun the neutrino depolarization
would be resonant and only in the solar convective zone one will find
a magnetic field $\vec{B}_\odot$ extended over such a wide region
($\sim$ a tenth of the solar radius). At earth-bound experiments the
helicity-flip could be caused only by the presence of a neutrino mass
and is therefore small.

Barbieri and Fiorentini considered \cite{Barbieri} the conversions
$\nu_{eL} \to \nu_{eR}$ as a result of the spin-flip by a
toroidal magnetic field in the solar convective zone. They showed that the
azimuthal asymmetry could be observable in a real time solar
$^8$B-neutrino experiment and as large as 20\% for an electron kinetic
energy threshold of $5$ MeV, fixing the $\nu_e$ survival
probability $P_e=1/3$ (as was suggested by the Homestake
experiment) and the maximal Dirac magnetic moment allowed by
laboratory experiments, $\mu_\nu\simeq 10^{-10}\mu_B$.  On the other
hand, Vogel and Engel \cite{Vogel} emphasized that if an asymmetry in
the scattering of transversally polarized neutrinos exists, recoil
electrons will be emitted copiously along the direction of the
neutrino polarization in the plane orthogonal to the neutrino
momentum.
However both \cite{Barbieri} and \cite{Vogel} {\em overestimated} the
asymmetry by an approximate factor two.  First, the weak term in
eq.~(11a) of ref. \cite{Barbieri} is a factor 2 less than our \eq{em}
and the interference term coincides with ours, while in ref.
\cite{Vogel} the interference term (their eq.~(A9)) is a factor 2
bigger than our \eq{diracdiag}.

Here we show the sensitivity of planned solar neutrino experiments in
the {\em low energy region} ($\omega \lsim 1$ MeV) to the azimuthal
asymmetries that are expected in the number of neutrino events,
arising from the electromagnetic-weak interference term. 

\begin{figure}[t]
\centerline{\protect\hbox{\psfig{file=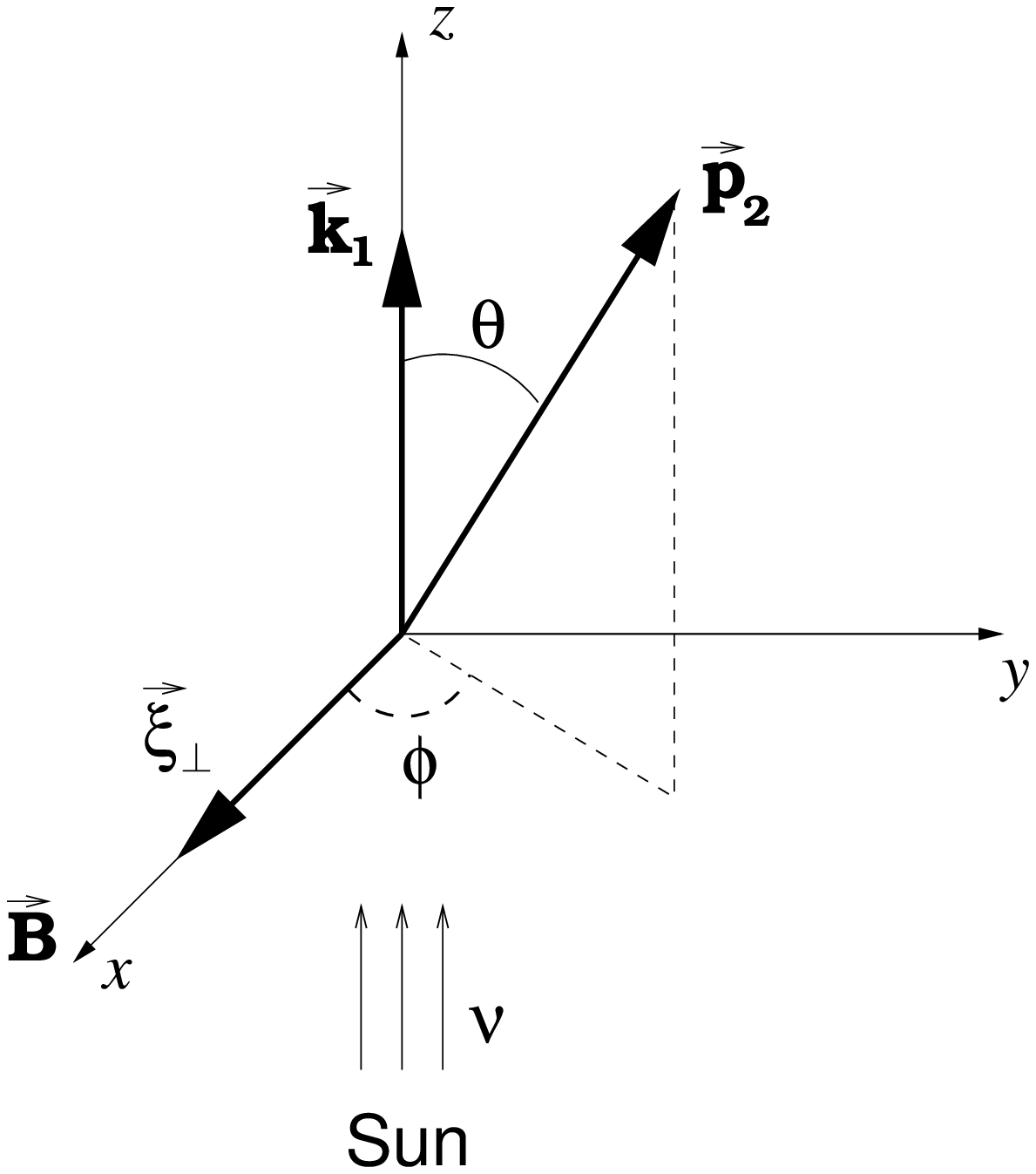,width=5cm}}}
\fcaption{Coordinate system conventions.}
\label{axis}
\end{figure}

We consider the neutrino--electron scattering process $\nu_e(k_1)
+e(p_1) \to \nu_e(k_2) +e(p_2)$ when the initial flux of neutrinos is
not completely polarized due to transitions induced by non-zero
transition magnetic moments in the Sun.  This includes both
conventional Dirac-type magnetic moments as well as Majorana
transition moments. The differential cross section $d\sigma/dTd\phi$,
where $T$ is the recoil energy of electrons and $\phi$ the azimuthal
angle (see fig. \ref{axis}),
can be written as a sum of three terms: weak, electromagnetic and
interference.
%
%
Let us first assume that only the Dirac $\nu_e$ magnetic moment
exists, $\mu_{\nu_e}$. For ultra-relativistic neutrinos the
expressions for the weak, electromagnetic and interference 
terms are 
\begin{equation}
\left (\frac{d\sigma}{dTd\phi}\right )_{weak} =
P_e\frac{G_F^2m_e}{\pi^2} h(g_{eL},g_R) \qquad
%
\left
 (\frac{d\sigma}{dTd\phi}\right )_{em} = \frac{\alpha^2 }{2m_e^2}
\left (\frac{\mu_{\nu_e}}{\mu_B}\right
)^2\left [\frac{1}{T} - \frac{1}{\omega}\right ]
\label{em}
\end{equation}
%
%
\begin{equation}
\left (\frac{d\sigma}{dTd\phi}\right )_{int} =
-\frac{\alpha G_F}{2\sqrt{2}\pi m_e T} 
\left(\frac{\mu_{\nu_e}}{\mu_B}\right ) \left[g_{eL} + 
g_R\left (1 - \frac{T}{\omega}\right )\right ]
\vec{p}_2 \cdot \vec{\xi}_\perp  
\label{diracdiag}
\end{equation}
where all parameters are the same as in \eq{totdsigma}.  The {\em
  interference term} is proportional to $\mu_\nu$ \cite{Barbieri}, and
$\vec{\xi}_\perp$ is the transverse component of the neutrino
polarization spin vector with respect to its momentum. It has the
value $\mid \vec{\xi}_{\perp} \mid = 2(P_e (1 - P_e))^{1/2}$. This
interference term depends on the azimuthal angle $\phi$ as defined in
figure \ref{axis}, since \eq{diracdiag} may be rewritten using
\begin{equation}
\vec{p}_2\cdot\vec{\xi}_\perp = \mid \vec{p}_2 \mid 
\sin \theta \mid \vec{\xi}_\perp \mid \cos \phi = \sqrt{2m_e
T\left (1 - \frac{T}{T_{max}}\right )} \mid \vec{\xi}_\perp \mid \cos
\phi
\label{angles}
\end{equation}
where $T_{max} = 2\omega^2/(m_e + 2\omega)$ is the maximum electron
recoil energy.

If neutrinos are Majorana particles they can only possess transition
magnetic moments, that we will denote $\mu_{12}$. For simplicity we
assume the case of CP conservation. For definiteness, moreover, we
consider the case of two neutrino species, $\nu_e$ and $\nu_\mu$, with
positive relative CP-parity \cite{BFD}.  The three terms of the
differential cross section will include an electromagnetic term (same
as in \eq{em} with $\mu \rightarrow \mu_{12}$), a weak term
that consists of the first and fourth terms in \eq{totdsigma},
%
%
and an interference term which differs from \eq{diracdiag},
\begin{equation}
-\frac{\alpha G_F}{4\sqrt{2}\pi m_e T} 
\left(\frac{\mu_{12}}{\mu_B}\right )
\left[(g_{eL} + g_{\mu L} + 2g_R)
\left (2 - \frac{T}{\omega}\right )+
(g_{eL} - g_{\mu L})
\frac{T}{\omega}\right ]
\vec{p}_2 \cdot
\vec{\xi}_\perp^{~e \bar{\mu}}
\label{majtran}
\end{equation}
since in the Majorana case there are two active neutrino species. Here
the mixed transversal polarization vector is given \cite{Sem97} by
$\mid \vec{\xi}_\perp^{~e\bar{\mu}} \mid = 2 \sqrt{P_e (1-P_e)}$.

The relevant quantity to be measured in neutrino-electron scattering
experiments capable (like HELLAZ) of measuring directionality of the
outgoing $e^{-}$ is the azimuthal distribution of the number of
events, namely
\begin{equation}
\label{spectrum2}
\frac{dN}{d\phi} = N_e \sum_{i} \Phi_{0i} 
\int^{T_{max}}_{T_{Th}}dT~
\int^{\omega_{max}}_{\omega_{min}(T)}
d \omega~ \lambda_i(\omega) 
\frac{d \sigma}{dTd\phi} (\omega,T) = n^{w}+n^{em}+
n^{int}\cos\phi
\end{equation}
where $d \sigma/dTd\phi$ is the total cross section and all parameters
are the same as in \eq{spectrum}. It has been written
as a sum of three terms, where $n^{w}$, $n^{em}$ and $n^{int}$
account for the weak, electromagnetic and interference contributions, 
respectively. The differential azimuthal asymmetry is defined as
\begin{equation}
\left.\Frac{dA}{d\phi}\right|_{\phi '}=
\Frac{\left.\Frac{dN}{d\phi}\right|_{\phi '}-
\left.\Frac{dN}{d\phi}\right|_{\phi '+\pi}}{
\left.\Frac{dN}{d\phi}\right|_{\phi '}+
\left.\Frac{dN}{d\phi}\right|_{\phi '+\pi}}=
\Frac{n^{int}}{n^{w}+n^{em}} \cos\phi '
\end{equation}
where $\phi '$ is measured with respect to the direction of
$\vec{B}_\odot$ , which we will assume to be along the positive
$x$-axis. One can also define an integrated asymmetry
\begin{equation}
{\cal A}(\phi ')=\Frac{\int^{\phi '+\pi}_{\phi '}\Frac{
dN}{d\phi}d\phi - \int^{\phi '+2\pi}_{\phi '+\pi} \Frac{
dN}{d\phi}d\phi}{\int^{\phi '+\pi}_{\phi '}\Frac{
dN}{d\phi}d\phi + \int^{\phi '+2\pi}_{\phi '+\pi} \Frac{
dN}{d\phi}d\phi}=-\Frac{2n^{int}}{\pi (n^{w}+n^{em})}\sin\phi '
\equiv -A \sin \phi '
\label{asym4}
\end{equation}
Here we have defined $A$, the maximum integrated asymmetry measurable
by the experiment, which is manifestly positive.

It is important to emphasize that HELLAZ will be the first experiment
potentially sensitive to azimuthal asymmetries since the
directionality of the outgoing $e^{-}$ can be measured. The angular
resolution is expected to be $\Delta \theta, \Delta \phi \sim 30$ mrad
$\sim 2^\circ$, substantially better than that of Super-Kamiokande.
Notice also that the Cerenkov cone defined by the angle $\theta$ is
very narrow for high-energy boron neutrinos, as one can see from
\eq{angles}.  In contrast, for $pp$ neutrino energies accessible at
HELLAZ ($T_{max} \simeq 0.26$ MeV, $T_{th} \simeq 0.1$ MeV) we
estimate that $\theta$ can be as large as $48^\circ$.

Let now discuss how the measurement of the azimuthal asymmetry could
be carried out considering that $\vec{B}_\odot$ is constant over a
given period of time but its direction is unknown. One should collect
events in every $\phi$-bin, where $\phi$ is defined with respect to
some arbitrarily chosen axis, and then take for different $\phi 's$
the ratio ${\cal A} (\phi ')$ which should show a $\sin \phi '$
dependence with a maximum equal to $A$.  This maximum will show us the
angle $\phi_{0}$ which corresponds to the direction of $\vec{B}_\odot$
($\phi_{0}=0$ if $\vec{B}_\odot$ goes along the positive x axis). Then
the direction of $\vec{B}_\odot$ is measured together with $A$. Since
this direction may change in time the experiment should accumulate
events until the maximum $\sin \phi $-like correlation in ${\cal A}
(\phi)$ is found and then start a new event counting period when such
correlation goes away due to the changing $\vec{B}_\odot$. Therefore
the value of $A$ could be extracted by performing a series of such
measurements.

\begin{figure}
\centerline{\protect\hbox{\psfig{file=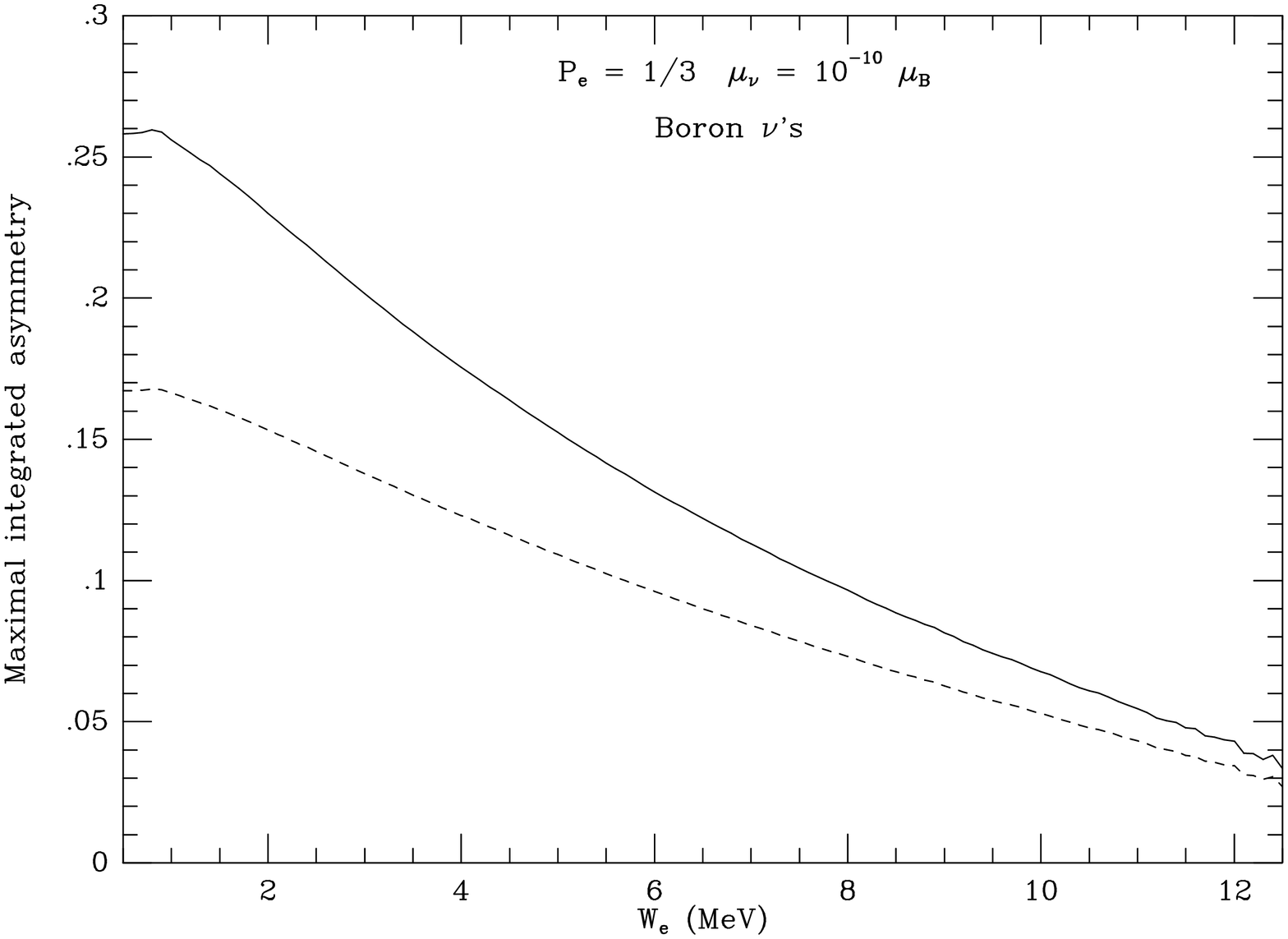,width=13cm,angle=-90}}}
\vspace{0.75cm}
%
\centerline{\protect\hbox{\psfig{file=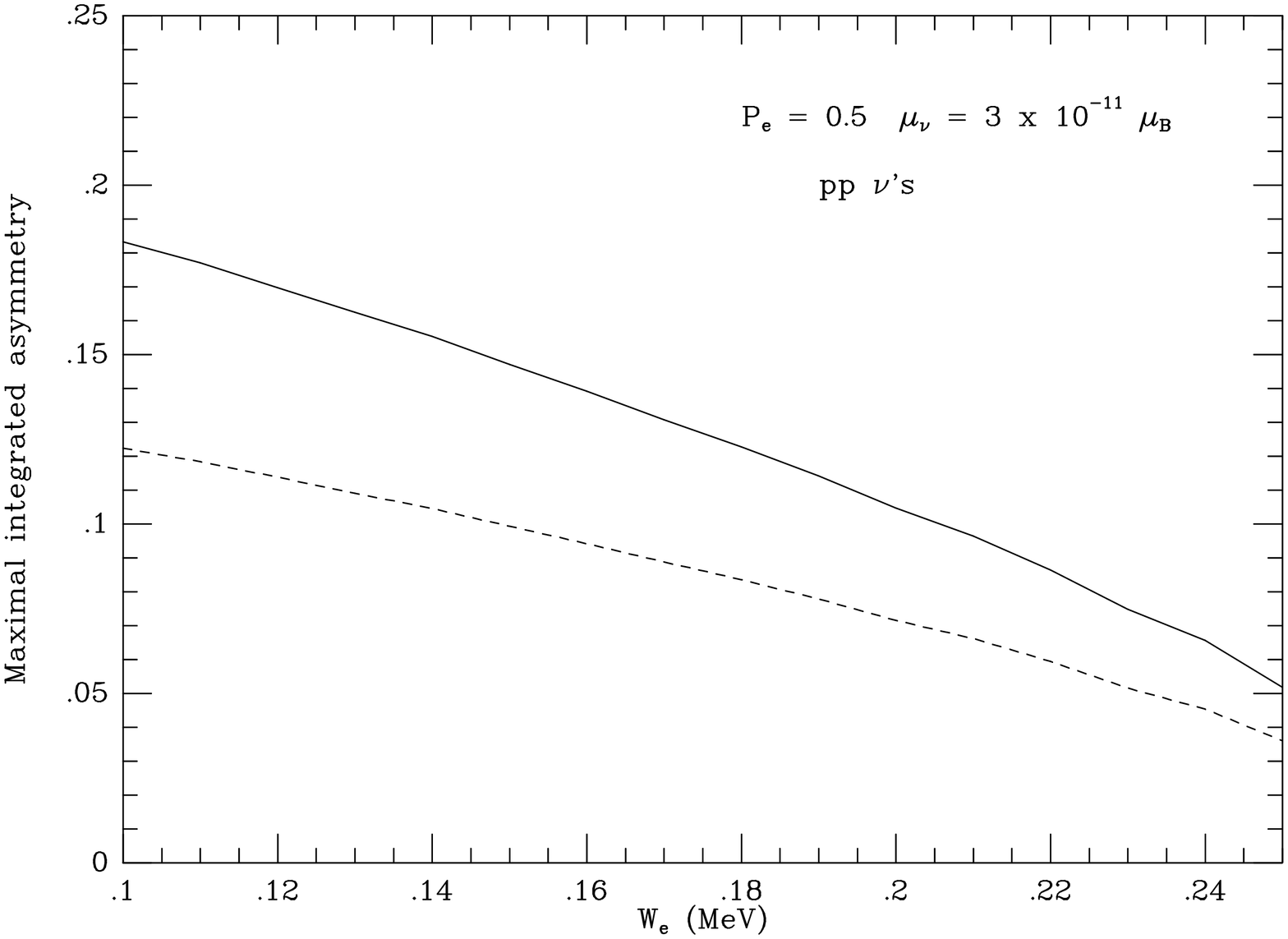,width=13cm,angle=-90}}}
\vspace{0.75cm}
\fcaption{Maximal integrated azimuthal asymmetry $A$ for Boron
neutrinos and $pp$ neutrinos, respectively, 
as a function of the electron recoil
energy threshold $W_e$ for the parameters shown. Solid line: Dirac
case ($\nu_{e L} \to \nu_{eR}$).  Dashed line: Majorana case
($\nu_{e } \to \bar{\nu}_{\mu }$)}
\label{asimlow}
\end{figure}
The first plot of fig. \ref{asimlow} corresponds to the values of $A$
in the situation described in refs. \cite{Barbieri,Vogel}. From our
calculations the asymmetry in the Dirac case is in fact
approximately a factor two smaller than predicted by \cite{Barbieri}.
The asymmetry in the Majorana case for equivalent $\mu$, as expected,
is smaller since there are two active neutrino species so
that the weak term in the denominator in \eq{asym4} becomes larger.
The second plot shows the same kind of analysis for $pp$-neutrinos,
which will be measured by an experiment like HELLAZ \cite{hellaz},
where both the recoil electron energy $T$ and the recoil electron
scattering angle $\theta$ should be measured with good precision. The
Multi-Wire-Chamber in HELLAZ should be sensitive to the
azimuthal angle $\phi$, measuring the number of events in
$\phi$--bins. We use a more realistic survival probability for 
Resonant Spin-Flavour Precession (RSFP) in
the Sun ($P_e=0.5$) and for $\mu_\nu = 3\times 10^{-11}\mu_B$. This
gives the maximum expected asymmetry and seems phenomenologically
reasonable in order to convert the initial solar $\nu_{eL}$'s via the
RSFP scenario.

The dependence of $A$ on the value of $\mu_\nu$ deserves a
more detailed analysis. It follows that $A$ is maximized when the pure
weak term is equal to the electromagnetic contribution. This fact
favours $pp$-neutrinos with respect to high energy neutrinos, since
such a maximum is reached for lower $\mu_{\nu}$ values precisely due
to the lower energies considered. In fact, for an energy threshold of
recoil electrons $W_{e}=0.1$ MeV (HELLAZ) the maximal asymmetry is
reached for $\mu_{\nu}\simeq 3 \times 10^{-11}\mu_{B}$ in contrast to
Boron neutrinos, for which it is reached for $\mu_{\nu} \simeq
10^{-10} \mu_{B}$. This way one sees that figs.~\ref{asimlow}
describe approximately the most favourable situation for
measuring $A$.

The fact that $dA/d\mu_{\nu}=0$ is reached for equal number of weak
and electromagnetic events seems to suggest that the measurement of
the total number of events would be enough to rule out the values of
$\mu_{\nu}$ to which the asymmetry is sensitive. However, it is not so
if one bears in mind that background events from other processes
always increase the total number of events and one needs to perform a
good background subtraction to get some information on the $\mu_{\nu}$
term. In addition if the background is isotropic in the azimuthal
plane, it should be absent in the numerator of $A$ and then if
some asymmetry is measured one can ascribe it to an effect of
$\mu_\nu$. Note also that the asymmetry is a ratio of event numbers.
Thus global normalization uncertainties (e.g.  total neutrino fluxes)
completely drop out from the asymmetry, and energy-dependent
uncertainties will be reduced since the same integrations appear in
the numerator and in the denominator. For all these reasons the
asymmetry measurement is preferred. It suffers of course from the
dependence on the unknown magnetic field direction, but sensitivity to
that information is also of potential astrophysical interest.

We have considered\cite{asim} how the above results are affected by
the energy dependence expected in the conversion probability in the
simplest realization of the RSFP scenario \cite{LAM} where vacuum
mixing is neglected.  It can be shown that the dependence of $P_e$ on
the neutrino energy leads to somewhat smaller azimuthal asymmetries,
but qualitatively very similar to those obtained if the energy
dependence is neglected.

We conclude that measuring azimuthal asymmetries in future low-energy
solar neutrino-electron scattering experiments with good angular
resolution should be a feasible and illuminating task. It
should provide useful information on non-standard neutrino
properties such as magnetic moments, as well as on solar magnetic
fields. 

Note we have assumed $\nu_e$ magnetic moments of the order
$10^{-11}\mu_B$, which is consistent with present laboratory
experiments and, apart from possible effects in red giants, a also
compatible with astrophysical limits \cite{RaffeltA}, given the
present uncertainties in these considerations. Finally, note that the
future ITEP-Minnesota expe\-riment is planned for searching
$\mu_{\nu}/\mu_B$ down to $3\times 10^{-11}$ with reactor
anti-neutrinos \cite{Voloshin}, while the LAMA experiment
\cite{Bernabei} will use a powerful isotope neutrino source.

\vspace*{0.25cm}
{\normalsize\bf Acknowledgements}
\vspace*{0.25cm}

Work supported by DGICYT under grant PB95-1077, 
by the TMR network grant ERBFMRXCT960090 and by INTAS grant
96-0659 of the European Union. S.P. was supported by Conselleria
d'Educaci\'o i Ci\`encia of Generalitat Valenciana. 


\vspace*{0.25cm}
{\normalsize\bf References}
\vspace*{0.25cm}

\end{document}